\documentclass[sts]{imsart}

\RequirePackage{amsthm,amsmath,amsfonts,amssymb,graphicx}
\RequirePackage[authoryear]{natbib}
\RequirePackage[colorlinks,citecolor=blue,urlcolor=blue]{hyperref}

\startlocaldefs

\usepackage{color}

\theoremstyle{plain}

\newcommand{\E}{\mbox{$\mathbb{E}$}}
\newcommand{\V}{\mbox{$\mathbb{V}$}}
\newcommand{\C}{\mbox{$\mathbb{C}$}}
\newcommand{\R}{\mbox{$\mathbb{R}$}}
\let\hat\widehat

\let\bar\overline

\definecolor{purple}{RGB}{150,0,100}

\definecolor{myblue}{RGB}{50,50,150}
\usepackage{sectsty}
\allsectionsfont{\color{myblue}\bfseries\sffamily}

\endlocaldefs

\begin{document}

\begin{frontmatter}
\title{\textcolor{myblue}{Feature Importance: 
A Closer Look at Shapley Values and LOCO\thanksref{T1}}}
\runtitle{Feature Importance}
\thankstext{T1}{The authors thank Art Owen for helpful comments.}

\begin{aug}
\author[A]{\fnms{Isabella} \snm{Verdinelli}
\ead[label=e1]{isabella@stat.cmu.edu}}
\and
\author[B]{\fnms{Larry} \snm{Wasserman}
\ead[label=e2]{larry@stat.cmu.edu}}

\address[A]{Isabella Verdinelli is Professor in Residence, 
Department of Statistics and Data Science, Carnegie Mellon University, 
Pittsburgh, PA USA, \printead{e1}.}
\address[B]{Larry Wasserman is Professor,
Department of Statistics and Data Science,
and Professor, Machine Learning Department,
Carnegie Mellon University, PA USA, \printead{e2}.}
\end{aug}

\begin{abstract}
There is much interest lately in explainability in
statistics and machine learning.
One aspect of explainability is to quantify the importance
of various features (or covariates).
Two popular methods for defining variable importance are
LOCO (Leave Out COvariates) and Shapley Values.
We take a look at the properties of these methods and their
advantages and disadvantages.
We are particularly interested in the effect of correlation between 
features which can obscure interpretability.
Contrary to some claims, Shapley values do not
eliminate feature correlation.
We critique the game theoretic axioms for Shapley values
and suggest some new axioms.
We propose new, more statistically oriented axioms for feature 
importance and some measures that satisfy these axioms.
However, correcting for correlation is a Faustian bargain: 
removing the effect of correlation creates other forms of bias.
Ultimately, we recommend a slightly modified version of LOCO.
We briefly consider how to modify Shapley values
to better address feature correlation.
\end{abstract}


\end{frontmatter}

\section{Introduction}

In this paper, we review
some notions of feature (covariate) importance in regression,
a topic that has received renewed interest lately.
There are many different ways of defining feature
importance.
Indeed, as \cite{hama2022} say:
\begin{quote}
{\em ... there is no ground truth for variable importance. 
Instead, there are multiple definitions of what makes a variable important, and choosing a definition involves some tradeoffs.}
\end{quote}
Population level feature importance is concerned with
measuring how important a particular feature is
in the underlying data generating process.
For example, how important is age in a regression function $\mu(x)$?
Algorithmic feature importance measures
the importance of a feature in a given estimator
without reference to an underlying distribution.
For example, how much did age affect the value $\hat\mu(x)$?
Causal feature importance
refers to counterfactual claims.
For example, how would the outcome have changed
if Mary had been 5 years younger?
We will focus on population level feature importance.
Another distinction is global importance versus local
importance. The former refers to the overall importance of a feature
while the latter refers to one point in the sample space.
This paper deals with global importance.

Even within the realm of population level feature importance
there are a number of ways to
define importance. 
We consider two specific measures: LOCO (Leave Out COvariates)
\citep{lei2018distribution, rinaldo2019bootstrapping,
williamson2021nonparametric, williamson2020unified,
gan2022inference}
and Shapley values \citep{shapley, lundberg2017unified, 
covert2020understanding, covert2021explaining, hama2022,
chen2022algorithms, kumar2020problems}.
Our main conclusion is that
Shapley values are much more complicated and expensive than LOCO
but have little to offer beyond LOCO.
In particular, both measures are compromised by
correlation between features which makes interpretation difficult.
There seems to be some claims in the literature that 
Shapley values reduce the effect of correlation 
(see \cite{iooss, williamson2020efficient} for example)
but we show that this is not true.
It is possible to reduce the effect of correlation
by modifying LOCO.
But this modification has a price:
the resulting estimator is subject to first-order estimation bias
(unlike LOCO) and, moreover, is subject to further bias
when there are regions of low density in the sample space.
Ultimately, there is no perfect measure
and we suggest using a minor variation of LOCO
as being the simplest and most reliable.

{\bf Related Work.}
The literature on feature importance
has exploded recently.
Some of the recent references on LOCO
include:
\cite{lei2018distribution, rinaldo2019bootstrapping,
williamson2021nonparametric, williamson2020unified,
gan2022inference}.
The literature on
Shapley values is vast and we cannot attempt
an exhaustive list here.
The references most related to
this paper include
\cite{shapley, lundberg2017unified, 
covert2020understanding, covert2021explaining, hama2022,
chen2022algorithms,hama2022,kumar2020problems, mase2022}.
However, this is just the tip of the iceberg.
Most papers on Shapley values for feature importance
appear in the AI and machine learning venues
as well as in the applied math literature that deals with
computer experiments.
As mentioned in the introduction,
there are many other feature importance measures
besides
LOCO and Shapley values.
These include
LIME \citep{ribeiro2016model},
Floodgate \citep{zhang2020floodgate}
and the somewhat related area of knockoffs
\citep{barber2015controlling}.

{\bf Paper Outline.}
In Section \ref{sec::locoandshapley}
we define LOCO and Shapley values.
In Section
\ref{sec::correlation}
we discuss the effect of feature correlation.
In Section \ref{sec::axioms}
we review the axioms that are the basis for Shapley values
and we consider alternative axioms.
In Section \ref{sec::corrfree}
we discuss methods for obtaining correlation free
measures of feature importance.
We discuss a variety of inferential issues 
in Section \ref{sec::inference}.
Section \ref{sec::examples} contains a few simulated examples.
Groups of features are discussed in Section
\ref{sec::groups}.
In Section \ref{sec::modified} we introduce the idea
of explicitly modifying Shapley values to address feature correlation.
Finally, Section \ref{sec::conclusion}
contains concluding remarks and recommendations.

\section{LOCO and Shapley}
\label{sec::locoandshapley}

Let $X\in\R^d$ be a vector of features
and let $Y\in \R$ be an outcome.
Let
$\mu(x) = \E[Y|X=x]$ be the regression function.
We are interested in quantifying the importance of a feature $X_j$.
We will use negative subscripts to mean that features have been dropped.
Specifically, let
$$
\mu_{-j}(x) = \E[Y|X_{-j}=x_{-j}]
$$ 
denote the regression
when $X_j$ is omitted
and more generally let
$$
\mu_{-S}(x) = \E[Y|X_{-S}=x_{-S}]
$$
denote the regression when the set of features
$X_S = (X_j:\ j\in S)$ is omitted
where $S\subset\{1,\ldots, d\}$.
Also, let
$$
\mu_S(x) = \E[Y| X_S = x_s]
$$
be the regression of $Y$ on
$X_S = (X_j:\ j\in S)$.
We let $\E$, $\V$ and $\C$
denote expectation, variance and covariance.

\subsection{Definitions}

The LOCO parameter for $X_j$ is defined by
\begin{align}\label{eq::loco}
\psi_{loco}(j)& \equiv \psi_{loco}(X_j) 
= \E[ (\mu(X) - \mu_{-j}(X))^2] \\
& = \E[(Y-\mu_{-j}(X))^2] - \E[(Y-\mu(X))^2]. \nonumber
\end{align}
LOCO can be interpreted as the $L_2$
difference between the full model and the model without $X_j$
and it can also be interpreted as the drop in prediction
accuracy when $X_j$ is omitted.
LOCO is related to two other quantities.
First,
$\psi_{loco}(j) = \sigma^2( R^2 - R^2_{-j})$
where
$\sigma^2 = \V[Y]$,
and
$$
R^2 = 1 - \frac{\E[ (Y-\mu(X))^2]}{\sigma^2},\ \ \ 
R^2_{-j} = 1 - \frac{\E[ (Y-\mu_{-j}(X))^2]}{\sigma^2}.
$$
Second, 
it is easy to check that
$\psi_{loco}(j) = \E\biggl[ \V[\mu(X)|X_{-j}]\biggr]$.
The latter expression is known as 
the upper Sobol' index
\citep{owen2017shapley, sobol1990, sobol1993}.

For a set of features $S$, LOCO is defined to be
\begin{equation}\label{eq::locoS}
\psi_{loco}(S) = \E[(\mu(X)-\mu_{-S}(X))^2].
\end{equation}
The values in (\ref{eq::loco}) and
(\ref{eq::locoS}) refer to the effect of dropping $X_j$ or $X_S$
from the full model.
We can also define LOCO when dropping $X_j$ from a submodel 
$M\subset\{1,\ldots, d\}$.
Specifically, define
\begin{equation}
\label{eq::AA}
\psi_{loco}(j,M) = \E\Bigl[(\E[Y|X_{M}]-\E[Y|X_{M \symbol{92} \{j\}}])^2 \Bigr].
\end{equation}
LOCO has been studied in 
\cite{lei2018distribution, rinaldo2019bootstrapping,
williamson2021nonparametric, williamson2020unified,
gan2022inference}.
It has several nice properties: it is easy to estimate, easy to interpret
and relates to the familiar $R^2$ from linear regression.
However, it has one important drawback:
its value depends not just on how strongly $\mu(X)$
depends on $X_j$, but also on the correlation between $X_j$ and 
the other features.
In the extreme case of perfect dependence, it is 0.
This could lead users to erroneously conclude that some variables are irrelevant
even when $\mu(X)$ strongly depends on $X_j$.
We refer to this as {\em correlation distortion}.

Next we define Shapley values.
First, we assign a value $V(S)$ to the submodel based on the subset of features $S$.
Then the Shapley value for $X_j$ is defined by
\begin{equation}\label{eq::shap}
\psi_{shap}(j) \equiv
\psi_{shap}(X_j)=\sum_{S\subset -\{j\}} w_S\ \ \Bigl( V( S \cup\{j\}) - V(S)\Bigr)
\end{equation}
where the sum is over all $2^{d-1}$ submodels that don't include $X_j$
and 
\begin{equation}
w_S = \frac{1}{d}{\binom{d-1}{|S|}}^{-1}.
\end{equation}
In this paper, we take $V(S)$
to be the prediction error
$V(S) = \E[ (Y-\mu_S(X))^2]$
in which case, we have
$$
\psi_{shap}(j) = \sum_S w_S \ \psi_{loco}(j,S).
$$
Thus, 
with this choice
of value function,
the Shapley value is just a weighted average of the LOCO values over submodels.
Of course, there are many other choices for the value function:
see \cite{chen2022algorithms} for a thorough discussion.
The Shapley value was defined by \cite{shapley}
in the context of cooperative game theory.
It is the unique function that satisfies a set of axioms.
We discuss these axioms in Section \ref{sec::axioms}.
The Shapley value can also be written as
$$
\frac{1}{d!}\sum_\pi \Biggl[V\Bigl(S(\pi,j)\Bigr) - 
V\Bigl(S(\pi,j)\cup\{X_j\}\Bigr)\Biggr]
$$
where the sum is over all permutations $\pi$
of $\{1,\ldots, d\}$
and $S(\pi,j)$ denotes all features that appear before $X_j$
in permutation $\pi$.

\subsection{The Linear Model}
\label{sec::linear}

An important special case
is the linear model
$\mu(x) = \sum_j \beta_j X_j$.
Then, it is easy to show that
$$
\psi_{loco}(j) = \beta_j^2 \ \E[(X_j-\nu(X_{-j}))^2]
$$
where $\nu(x_{-j}) = \E[X_j|X_{-j}=x_{-j}]$.

\cite{owen2017shapley}
showed that, assuming that $X$ is Gaussian, the Shapley value can be written as
$$
\psi_{shap}(j) = \frac{1}{d}\sum_S
\binom{d-1}{|S|}^{-1} \frac{ \C[X_j, X_{-S}^T\beta_{-S}|X_S]^2}{\V[X_j|X_S]},
$$
where $\C[X,Y]$ denotes the covariance between random variables
$X$ and $Y$.
To get more insight,
we look at a particular linear model
endowed with some symmetries.
Let $X = (W,Z)$, 
where $W$ is the feature of interest and let
$$
Y = \beta W + \gamma Z_1 + \cdots + \gamma Z_{d} + \epsilon
$$
where $\epsilon$ has mean 0.
Assume that the $Z_i'$s
are exchangeable and define
$$
b_j \equiv b_j(W,Z_1, \ldots, Z_j) = \E[Z_{\bullet}|W,Z_1,\ldots, Z_j]
$$
where $Z_\bullet$ denotes
any other $Z_s$ with $s>j$.
Note that
$b_0 = \E[Z_\bullet|W]$.
After some algebra, the exchangeability of the $Z_i'$s leads to
\begin{align*}
\psi_{shap}(W) &=
\frac{\beta^2}{d+1}\V[W] +
\frac{\gamma^2 d^2}{d+1}\V[b_0] +
\frac{2\beta\gamma d}{d+1} \C[W,b_0]\\
& +
\frac{\beta^2}{d+1} \sum_{j=1}^d \E\bigl[\V[W|Z_1,\ldots,Z_j]\bigl] \\
& +
\frac{\gamma^2}{d+1} \sum_{j=1}^d (d-j)^2 \E[\V[b_j|Z_1,\ldots,Z_j]]\\
& +
\frac{2\beta\gamma}{d+1} \sum_{j=1}^d (d-j) \E[\C[W,b_j|Z_1,\ldots,Z_j]].
\end{align*}
Even in this simple case
we see that the Shapley value is a complicated
function of the joint distribution.

\subsection{Estimation}

Given data
$(X_1,Y_1),\ldots, (X_n,Y_n)$,
let $\hat\mu(x), \, \hat\mu_{-j}(x)$ and
$\hat\mu_{-S}(x)$ be estimates of
$\mu(x),\mu_{-j}(x)$ and $\mu_{-S}(x)$.
The plugin estimate of LOCO for $X_j$ is
$$
\hat\psi_{loco}(j) = \frac{1}{n}\sum_{i=1}^n (\hat\mu(X_i)-\hat\mu_{-j}(X_i))^2.
$$
The plugin estimate of the Shapley value
is
$$
\hat\psi_{shap}(j) = \sum_{S\subset-\{j\}} w_S \ \hat\psi_{loco}(j,S).
$$
Summing over all $2^{d-1}$ subsets is not practical.
Instead, $\psi_{shap}(j)$ is usually estimated by randomly sampling submodels
\citep{williamson2020efficient, covert2020understanding}.
Specifically, we can estimate
the Shapley values $\psi_{shap}(j)$
by sampling submodels
$S_1,\ldots, S_N$
with probabilities $w_S$
and letting
$$
\hat\psi_{shap}(j) = \frac{1}{N}\sum_r \hat\psi_{loco}(j,S_r).
$$
\cite{williamson2020efficient}
showed that, to get accurate estimates,
one needs to take $N = cn$ for some $c>0$.
This could mean refitting the model
thousands of times when the sample size is large.
The computational burden should not be underestimated.
This only estimates the Shapley value for $X_j$.
More generally,
one can get all the Shapley values
by doing a weighted regression as described in
\cite{williamson2020efficient,covert2020understanding}.
Specifically, the Shapley values can be obtained
by minimizing
\begin{equation}
\sum_S \frac{d-1}{\binom{d}{|S|}|S|(d-|S|)}
\Biggl[\sum_{i\in S}\psi_{shap}(i) - V(S)\Biggr]^2
\end{equation}
which is a weighted least squares problem.
This formulation also
shows that the Shapley values can be interpreted
as an additive approximation to the value function $V(S)$
albeit with non-intuitive weights.
In other words, $V(S)$ is
approximated by $\sum_{j\in S} \psi_{shap}(j)$.

\section{Feature Correlation}
\label{sec::correlation}

Correlation between features can cause 
difficulties with interpretation for both LOCO and Shapley.
For example,
consider the linear model
$\mu(x) = \sum_j \beta_j x_j$.
Then
$$
\psi_{loco}(1)\quad = \underbrace{\beta_1^2}_{\text{intrinsic importance}}
\ \underbrace{\E[ (X_1 - \nu_1(X))^2]}_{\text{feature correlation}}
$$
where
$\nu_1(X) = \E[X_1| X_2,\ldots,X_d]$.
Hence, the LOCO value is the product of
the intrinsic importance of $X_1$, given by $\beta_1^2$,
and the term 
$\E[ (X_1 - \nu(X_1))^2]$
which measures the correlation between $X_1$ and the other features.
In general, feature correlation
can decrease or increase the LOCO value.
The fact that LOCO depends on feature correlations
is technically correct.
After all, if $X_1$ is perfectly correlated with another feature,
then we can drop $X_1$ with no loss in prediction accuracy and LOCO would be 0
which is correct.
But, it seems likely that a user could misinterpret
this result and conclude that $\mu(x)$ does not depend on $X_1$,
even though $\beta_1$ could be large.

There seem to be some suggestions in the literature that
the Shapley value overcomes this problem.
For example,
\cite{iooss} say
\begin{quote}
{\em The Shapley effects were recently introduced to overcome this problem 
[correlation] as they allocate the mutual contribution (due to correlation and interaction) of a group of inputs to each individual input within the group.}
\end{quote}
They are correct that Shapley distributes the signal among
correlated features.
Whether this solves the problems due to correlation is
debatable.
Similarly, 
\cite{williamson2020efficient} writes:

\begin{quote}
{\em
Because [Shapley values] satisfy these properties, 
they clearly address the issue of correlated features: given collinear 
variables $X_j$ and $X_k$ that are each marginally predictive, previous 
nonparametric population 
[variable importance measures]
would assign zero importance to both variables 
whereas [Shapley] would assign the same positive value to both variables.
}
\end{quote}
This is true, but again it is not clear if this really solves the problem.
To illustrate this issue in more detail,
consider a simple example.
Suppose that
$$
Y = \beta X_1  + \gamma X_2 + \gamma X_3 + \cdots + \gamma X_d + \epsilon.
$$
We want to quantify the importance of $X_1$.
Assume that $\V[X_1]=1$.
We will look at $\psi_{loco}(1)$ and $\psi_{shap}(1)$.
Consider the following four cases:

{\bf Case 1.}
Suppose that $\beta \neq 0$, $\gamma = 0$
and the covariates are independent.
In this case we find that
$\psi_{loco}=\psi_{shap} = \beta^2$
so all the methods correctly find that $X_1$ is an important variable.

{\bf Case 2.}
Suppose that $\beta \neq 0$, $\gamma = 0$
but now suppose that the covariates are perfectly dependent, that is,
$P(X_1=X_2=\cdots = X_d) = 1$.
Now we get
$\psi_{loco}=0$
and $\psi_{shap} = \beta^2/d$
which is close to 0 when $d$ is large.
In this case, LOCO and Shapley are both subject to correlation distortion
and mislead us into thinking that $X_1$ is not important.

{\bf Case 3.}
Suppose that $\beta = 0$, $\gamma \neq 0$
and the covariates are independent.
Then
$\psi_{loco}= \psi_{shap} = 0$.
So both methods recognize that $X_1$ is not important.

{\bf Case 4.}
This is the same as Case 3 except that the covariates
are perfectly dependent.
Then
$\psi_{loco}=  0$
but
$\psi_{shap} = \gamma^2 d^2$.
Here the Shapley value can be large even though $Y$ has no
functional relationship with $X_1$.

\smallskip

This simple example
illustrates that
both $\psi_{loco}$ and $\psi_{shap}$
can be quite misleading in the presence of correlation.

\section{Axioms for Feature Importance}
\label{sec::axioms}

One reason why Shapley values are gaining popularity
is their axiomatic basis. The axioms are cited in most
papers that use Shapley values.
Here we review and critique these axioms
and we suggest some new axioms.

\subsection{The Shapley Axioms}

Let $S$ denote a subset of $\{1,\ldots, d\}$
and let ${\cal S}$ denote all such subsets.
Let $V:{\cal S}\to \mathbb{R}$
be a function that measures the value of the submodel
$X_S = (X_j:\ j\in S)$.
In our case, we use
$V(S) = \E[(Y - \mu(X_S))^2]$
where $\mu(X_S) = \E[Y|X_S]$.
Let $\psi(j)$ be the Shapley value for $X_j$.
The Shapley axioms are:

(S1: Efficiency) $\sum_j \psi(j) = V(\{1,\ldots, d\})$.

(S2: Symmetry) If $V(S \bigcup \{i\})= V(S \bigcup \{j\})$ for all $S$
then $\psi(i) = \psi(j)$.

(S3: Dummy) If $V(S\bigcup\{i\}) = V(S)$ for all $S$ then $\psi(i)=0$.

(S4: Linearity) 
If value functions $V$ and $V'$ yield Shapley values
$\psi(j)$ and $\psi(j')$ then
the value function $V+V'$ yields Shapley values 
$\psi(j)+\psi(j')$.

Shapley (1953) proved that the unique $\psi(j)$ that satisfies these axioms is
(\ref{eq::shap}).

\bigskip

{\bf Critique of The Shapley Axioms.}
Now we discuss some problems with Shapley values
and their axioms.

\begin{enumerate}

\item The Shapley axioms were intended to define the contribution
of each player in a cooperative game.
But when applied to feature importance in statistical models,
it is not clear why the axioms should refer to all submodels $S$.
Most submodels will be poor models with large predictive risk.
Why should we care how $\psi$ behaves on these submodels?

\item Axiom S1 requires additivity. But it is not at all clear
why we would expect feature importance to be additive.
For example, in the presence of interactions
we might expect feature importance to be non-additive.

\item The Shapley value is not interpretable.
What does a Shapley value of 7.2 mean?
Informally, users regard larger Shapley values to mean that one variable
is more important than another.
But if $\psi_{shap}(2) = 10 \psi_{shap}(1)$ does this mean that
$X_2$ is 10 times more important than $X_1$?
In contrast, LOCO has a clear meaning:
it is the increase in predictive risk when we drop $X_j$.

\item As we have seen, Shapley values suffer
from correlation distortion just as LOCO does.

\item Shapley values are expensive to compute as they involve
sums over large numbers of submodels.
Recall that \cite{williamson2020efficient} showed that
we need to subsample $O(n)$ submodels for accurate inference.
\end{enumerate}

\subsection{Alternative Axioms}

We view the correlation distortion
as a problem when trying to construct
interpretable measures of feature importance.
With this in mind,
we propose a
second set of axioms.
Write
$\psi(j) \equiv \psi(j,P)$ to denote a functional which measures
the importance of $X_S$
where $P$ is the distribution of
$(X,Y)$ which has density $p(x,y)$.

(A1: Functional Dependence) $\psi(j,P)=0$ if and only if
$\mu(x)$ is not a function of $x_j$.

(A2: Correlation Free) 
$$
\psi(j,P) = \psi\Bigl(p_{Y|X}(y|x), p_j(x_j),p_{-j}(x_{-j})\Bigr)
$$
where $p_j$ is the marginal of $X_j$ and
$p_{-j}$ is the marginal of $X_{-j}$.
That is, $\psi$ depends on $p_{Y|X}(y|x)$
and can depend on $p_X(x)$ only through the marginals
$p_j$ and $p_{-j}$
but not their joint distribution.

In a linear model, there is a natural
feature importance measure that is
automatically correlation-free, namely,
the regression coefficient.
With this in mind, our third axiom is:

(A3: Linear Agreement)
If $\mu(x) = \sum_j \beta_j x_j$ 
then $\psi(j,P) = \beta_j^2$.

\bigskip

A1 captures the idea that $\psi$ measures how strongly $\mu$ depends on $X_j$.

A2 expresses the fact that we want $\psi$ to measure how $Y$
is affected by $X_j$ but should be insensitive to the relationship between $X_j$ and 
$X_{-j}$.

A3 deserves some discussion.
It is our view that in a linear model,
the regression coefficient provides a clear, unambiguous
measure of feature importance.
Indeed, some approaches to local feature importance
explicitly fit a local linear model and use the
regression coefficient as a feature importance parameter
\citep{ribeiro2016model}.
We acknowledge that this point is debatable
but we use A3 in what follows.
However, A3 is not crucial and could be dropped 
without changing the main messages.

In general,
neither Shapley nor LOCO satisfy A2.
In fact, Shapley also violates A1.
However, we can make LOCO satisfy A3 by
re-defining LOCO as
\begin{equation}\label{eq::loconew}
\psi_{loco}(j) = 
\frac{\E[ (\mu_{-j}(X) - \mu(X))^2]}{\E[ (X_j - \nu_j(X)]^2}
\end{equation}
where $\nu_j(x) = \E[X_j|X_{-j}=x_{-j}]$.
This normalized version of LOCO is the version we will use from now on.
Interestingly, this version often suffers less
correlation distortion even in nonlinear models, although it does not eliminate it in general.

There are a few more
desiderata that we would
like $\psi$ to satisfy. 
They are too vague to be called axioms but are important.

(V1) $\psi$ should be easy to compute.

(V2) The plug-in estimate should have second order bias.
This point is explained in more detail in Section 
\ref{sec::inference}.

(V3) The estimate of $\psi$ should not incur large bias
due to low density regions in the sample space.

As mentioned earlier,
Williamson (2022) showed that
to accurately estimate Shapley values, we must subsample
order $n$ submodels which is quite expensive
and so violates V1.
We will shortly introduce correlation-corrected versions of
LOCO but these will, unfortunately, fail V2 and potentially V3.
The effect of low density regions
has been pointed by \cite{owen2017shapley} and
\cite{hooker2021}
and we discuss this 
in Section~\ref{sec::inference}.

\section{Correlation Free Feature Importance}
\label{sec::corrfree}

In this section, we define some functionals that satisfy
A1, A2 and A3.
For notational simplicity,
define
$X = (W,Z)$
where $W\in\R$ is the variable of interest
and $Z$ are the remaining features.

Our first example of a functional 
that satisfies A1-A3 is the decorrelated LOCO from
\cite{verdinelli2021decorrelated}.
This is based on the counterfactual question:
what value would LOCO have taken if $W$ and $Z$
had been independent?
To answer this question, let
$p^*(w,z,y) = p(y|w,z) p(w)p(z)$.
This is the closest distribution to $p$ 
in Kullback-Leibler distance that makes $W$ and $Z$
independent.
We define the decorrelated LOCO
$\psi_{Dloco}(W)$ to be the value LOCO takes
under $p^*$.
A simple calculation shows that
\begin{equation}
\psi_{Dloco}(W) = \frac{\int\int (\mu(w,z) -\mu_0(z))^2 p(w) p(z) dw dz}{\V[W]}
\end{equation}
where
$\mu_0(z) = \int \mu(w,z) p(w)dw$.
For the linear model
$\mu(w,z) = \beta w + \sum_j \beta_j z_j$
we see that
$\psi_{Dloco}(W) = \beta^2$.
We estimate $\psi_{Dloco}$
using the $U$-statistic
$$
\hat\psi_{0} =
\frac{1}{\binom{n}{2}}\
\frac{\sum_{i<j} (\hat\mu(W_i,Z_j) - \hat\mu_0(Z_j))^2}
{\hat\sigma^2_W}
$$
where
$\hat\mu_0(z) = \frac{1}{n}\sum_i \hat\mu(W_i,z).$

$\psi_{Dloco}$ will be our main example
of a correlation free parameter
but we also define two more decorrelated parameters
to show other possibilities.
Our second example is the average gradient
$$
\psi_{grad} =
\int\int (\mu'(w,z))^2 p(w)p(z) dw dz
$$
where 
$\mu'(w,z) = \partial \mu(w,z)/\partial w$.
Again, it is easy to check that
this satisfies (D1), (D2) and (D3).
This functional is related to
the definition in Samarov (1993)
who uses
$\int\int (\partial \mu(w,z)/\partial w)^2 dP(w,z).$

Our third functional, which is related to the controlled direct effect
$\mu(w_1,z) - \mu(w_2,z)$ from causal inference
\citep{robins1992}, is
$$
\psi_{cde} = 
\frac{\int\int\int \delta(w_1,w_2,z) p(w_1)p(w_2)p(z) dz dw_1 dw_2}{2\V[W]}
$$
where
$\delta(w_1,w_2,z) = (\mu(w_1,z)-\mu(w_2,z))^2$.

Having defined these functionals,
let us return to the linear example in Section \ref{sec::linear}.
Recall that the model is
$Y = \beta W + \gamma Z_1 + \gamma Z_2 + \cdots \gamma Z_d + \epsilon$
and that $\beta=0$ in Cases 3 and 4.
The results are summarized in Table \ref{table::simple}.
Keep in mind that LOCO is now the normalized version.
Because this is a linear model, $\psi_{loco}=\psi_{Dloco}$
and both behave well.
But more generally, they are not equal and $\psi_{loco}$
still suffers correlation distortion.
The Shapley value can be smaller or larger than
$\psi_{Dloco}$.

\begin{table}
\begin{center}
\begin{tabular}{cccc}
       & $\psi_{Dloco}$  & $\psi_{loco}$  & $\psi_{shap}$\\ \hline
Case 1 & $\beta^2$ & $\beta^2$ & $\beta^2$\\
Case 2 & $\beta^2$ & $\beta^2$       & $\beta^2/d$\\
Case 3 & $0$       & $0$       & $0$\\
Case 4 & $0$       & 0         & $\gamma^2 d^2$ \\
\\
\end{tabular}
\caption{\it \qquad The simple linear example.
We compare Decorrelated LOCO, LOCO and Shapley.}
\vspace{-.3in}
\label{table::simple}
\end{center}
\end{table}

Case 2 is especially important.
In this case,
$\psi_{Dloco}(W) =\beta^2$ and
$\psi_{shap}(W) = \beta^2/d$.
Because of the perfect correlation,
any confidence interval for $\psi_{Dloco}$ would have
infinite length.
For Shapley
we would conclude that
none of the variables are important which is misleading.
For $\psi_{Dloco}$ we would conclude that the importance of $W$
is not identified.
Which answer is considered correct is a matter of opinion.
Our view is that it makes sense to say that
the importance of $W$ is not estimable when $W$
is perfectly correlated with the other features.
For LOCO and Shapley,
the correlation changes the estimand itself.
For $\psi_{Dloco}$, the estimand is unaffected by
correlation but it is not identified.

\section{Inferential Issues}
\label{sec::inference}

We now turn to inference.
Under some smoothness conditions,
constructing asymptotic confidence intervals for
LOCO and Shapley is straightforward
\citep{williamson2020efficient,williamson2021nonparametric, williamson2020unified,
lei2018distribution,rinaldo2019bootstrapping}.
However, as we explain in this section,
when we use the same algorithm to construct an interval
for the decorrelated LOCO,
there will be a first order bias and the
interval will not have correct coverage.
It can still be interpreted as a variability interval, that is,
an interval which covers the mean of
the estimator $\bar{\psi}_n = \E[\hat\psi_n]$
rather than the parameter itself.
The first order bias can, in principle be removed
by computing the influence function but,
as we explain below, this is not practical.
Also, the decorrelated LOCO suffers from
extra bias when there are regions of low density.
LOCO and Shapley suffer from some problems due to the quadratic
nature of the parameter.
Here are the details.

\subsection{The Effect of Nuisance Functions}

A variable importance measure is a functional $\psi$ that
depends on nuisance functions.
For example, LOCO in (\ref{eq::loco}) depends on the nuisance functions $\mu$
and $\mu_{-j}$.
In general, to construct a confidence interval
for a parameter $\psi$,
we would like
$\sqrt{n}(\hat\psi - \psi)$
to converge to a mean $0$ Normal distribution.
Now
$$
\sqrt{n}(\hat\psi - \psi) = 
\sqrt{n}(\hat\psi - \bar{\psi}_n) + \sqrt{n}(\bar{\psi}_n - \psi)
$$
where $\bar{\psi}_n = \E[\hat\psi_n]$.
The term
$\sqrt{n}(\bar{\psi}_n - \psi)$
explodes if the bias is first order meaning that
$\bar{\psi}_n - \psi = O(||\hat\mu - \mu||)$.
For this reason, we would like the bias to be second order
meaning that
$\bar{\psi}_n - \psi = O(||\hat\mu - \mu||^2)$.
For example, if $\mu$ is in a Holder space of smoothness $\tau$
then $||\hat\mu-\mu|| = O_P(n^{-\tau/(2\tau+d)})$.
Then $\sqrt{n}||\hat\mu-\mu||\to \infty$ but
$\sqrt{n}||\hat\mu-\mu||^2\to 0$ if $\tau > d/2$.
Second order bias is usually achieved
by correcting the plug-in estimator
with the efficient influence function.
LOCO and Shapley already have second order bias
$||\hat \mu - \mu||^2$
without correction by the influence function
as long as $\mu(x)$ satisfies a strong smoothness condition.
This is shown in 
\cite{williamson2020unified}.
The decorrelated parameter $\psi_{Dloco}$ loses this property.
In principle we can correct this bias using the influence function.
But estimating the influence function turns out to be
difficult. For example,
it requires density estimation.
Moreover, the influence function is poorly estimated
especially when there is strong correlation, precisely when
$\psi_{Dloco}$ would be most useful and this method turns out to be impractical. 

A problem with $\psi_{Dloco}$,
and any correlation free measure,
is that there is bias due to
regions of low density in the sample space.
More precisely,
we need to estimate the regression function over regions
where the density is small and this results
in extrapolation bias.
This problem occurs for many types of feature importance.
\cite{hooker2021}
show that this occurs for
Breiman's permutation measure for random forests.
\cite{owen2017shapley}
call this the {\em data hole problem}.
See also \cite{mase2022} and \cite{kumar2020problems}.

\begin{figure}
\begin{center}
\vspace{-.4in}
\includegraphics[scale=.5]{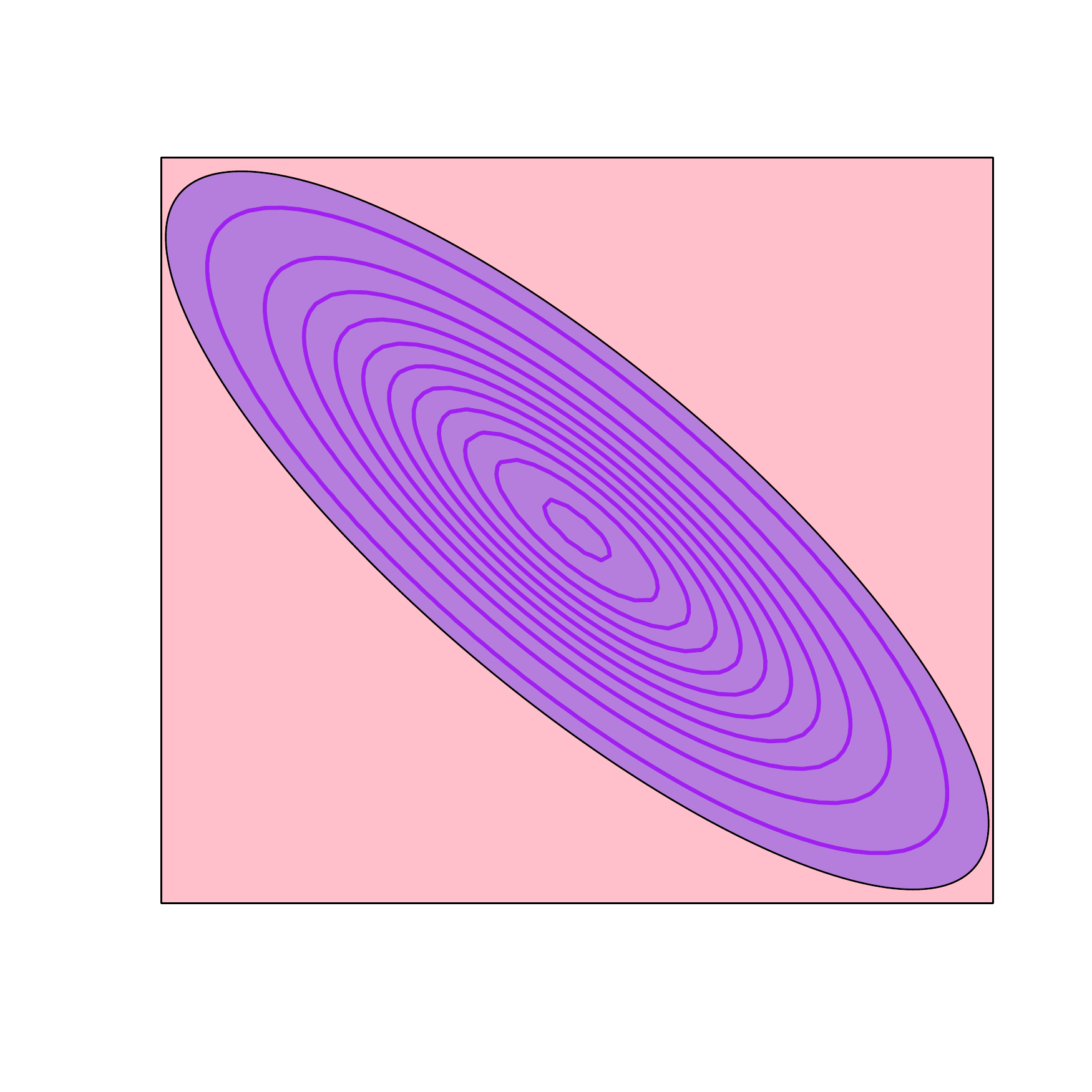}
\vspace{-.8in}
\caption{\it  The elliptical purple area is the bulk of the support
of the distribution. The square pink area is the region where
$p(w,z)$ is small but $p(w)p(z)$ is large.
There is little data in this region but any correlation free
functional depends on the value of the regression function 
$\mu(w,z)$ over this region.}
\label{fig::hole}
\end{center}
\vspace{-.1in}
\end{figure}

LOCO avoids this bias
so there is a tradeoff: correcting for correlation
eliminates correlation distortion at the expense
of introducing a new bias.
To see the problem,
note that 
$\psi_{Dloco}$ has the form
$\psi = \int\int g(w,z) p(w)p(z) dw dz$.
For any fixed value of $z$,
$\int\int g(w,z) p(w)p(z) dw dz$ requires integrating over all
values of $w$.
Thus we can end up integrating over pairs $(w,z)$
that have very low density; see Figure \ref{fig::hole}.
In particular, the estimator involves the sum
$\sum_{i,j} \hat\mu(W_i,Z_j)$
which includes pairs $(W_i,Z_j)$ which might lie in regions
with no observations.
In this case, $\hat\mu(W_i,Z_j)$ will either have
large bias or large variance.
For example, a random forest will extrapolate over this type of region
and have huge bias.
LOCO avoids this problem as it integrates
with respect to the joint distribution
$p(w,z)$.
To summarize, any nonparametric, correlation free
functional requires
estimating $\mu(w,z)$ over sets
of low density
which contain little data.
This raises the following question: what is worse:
Correlation distortion or low density bias?
We investigate this question in Section 
\ref{sec::examples}
where we consider some examples.

\subsection{Problems due to Quadratic Functionals}

Another issue that arises is that all these parameters are 
quadratic.
It is difficult to get valid confidence intervals 
or even variability intervals for
quadratic parameters because their
limiting distribution and rate of convergence change
as $\psi$ approaches 0.
Formally, the influence function vanishes when $\psi = 0$
and the central limit theorem does not hold.

With sufficient smoothness assumptions,
the estimator $\hat\psi$ of any of these parameters
converges at rate $n^{-1/2}$ 
when $\psi \neq 0$
but at the null, where $\psi=0$,
the influence function for the parameter vanishes,
the rate becomes $n^{-1}$
and the limiting distribution is typically
a combination of $\chi^2$ random variables.
Near the null, we get behavior in between these two cases.
A valid confidence interval $C_n$ should satisfy
$P(\psi_n \in C_n)\to 1-\alpha$
even if $\psi_n$ is allowed to change with $n$.
In particular, we want to allow $\psi_n\to 0$.
Finding a confidence interval with 
uniformly correct coverage,
with length $n^{-1/2}$ away from the null
and length $n^{-1}$ at the null
is, to the best of our knowledge, an unsolved problem.

We are aware of only three proposals
for handling this problem.
\cite{williamson2020unified}
deal with this problem when estimating $\psi_{loco}$ by
writing $\psi_{loco}$ as a sum of two
parameters
$\psi_{loco} = \psi_1 + \psi_2$
such that neither $\psi_1$ nor $\psi_2$ vanish
when $\psi_{loco}=0$.
Then, they estimate $\psi_1$ and $\psi_2$
on separate splits of the data.
This amounts to adding noise of size $O(1/\sqrt{n})$
to the interval.
In fact there are infinitely many ways to decompose $\psi$
as a sum of two non-degenerate functionals and this is just one.

\cite{shen}
instead add noise of the form $c Z/\sqrt{n}$ 
to the estimator,
where $Z\sim N(0,1)$. 
They choose
$c$ by permuting the data many times
and finding a $c$ that gives good coverage under the simulated permutations.
However, this is computationally expensive
and adding noise seems unnecessary.

Both of these approaches have the effect of
expanding the interval by an amount of order $O(n^{-1/2})$.
The proposal in
\cite{verdinelli2021decorrelated}
is simply to expand the interval
without adding any additional randomness.
Given an initial interval
$C_n =[a_n,b_n]$,
we expand it as
$C_n = [a_n - s/\sqrt{n},b_n + s/\sqrt{n}]$
where $s^2 = n^{-1}\sum_i (Y_i-\bar{Y}_n)^2$.
Any expansion of the form $c/\sqrt{n}$ would work;
we choose $c=s$ to set the scale to the scale of $Y$.
However, the method is quite insensitive to the choice of $c$.

All three methods are ad-hoc
but we are not aware of any other way to deal with the problem.
To see why these strategies
lead to valid coverage,
consider an estimator $\hat\psi_n$
with the following behavior.
When $\psi_n$ stays bounded away from 0,
we have 
$\sqrt{n}(\hat\psi_n - \bar{\psi}_n)/se_n\rightsquigarrow N(0,1)$
where $se_n$ is the estimated standard error
and $\bar{\psi}_n = \E[\hat{\psi}_n]$.
But when $\psi_n\to 0$,
$\V[\hat\psi_n] \asymp n^{-\gamma}$
with $\gamma>1$.
Let
$C_n = [\hat \psi_n - z_{\alpha/2} se_n - c/\sqrt{n},\ 
\hat \psi_n + z_{\alpha/2} se_n + c/\sqrt{n}]$.
When $\psi_n$ stays bounded away from 0,
the coverage is $\geq 1-\alpha$ by the central limit theorem.
When $\psi_n\to 0$,
the coverage is $\geq 1-\alpha$ by Markov's inequality.
All three approaches are basically the same:
they expand
the confidence interval 
by $O(n^{-1/2})$
which maintains validity at the expense
of efficiency at the null.

\subsection{Confidence Intervals}

\begin{table}
{\it Illustration of Cross-Fitting}
\begin{center}
\fbox{\parbox{3.6in}{
\begin{center}
\begin{tabular}{lc}
$D_1$ & \hspace{-3em} $\hat{\mu}_1$ \qquad from \qquad $D_2\ldots D_K$\ ,\\
      & \hspace{-6em} $\hat{\psi}_1$ \qquad from \qquad $D_1$\ .\\
\\
$D_2$ & \hspace{-1.8em}$\hat{\mu}_2$ \qquad  from \quad
                         $D_1,D_3,\ldots D_K$\ , \\
      & \hspace{-9.2em} \qquad $\hat{\psi}_2$ \qquad from \quad $D_2$\ .\\
$\vdots$      &                   $\vdots$    \\
$D_j$ &\hspace{3.5em} $\hat{\mu}_j$ \qquad from 
                         \quad $ D_1,\ldots,D_{j-1},D_{j+1},
                                     \ldots,D_K$ \\
      & \hspace{-8em} $\hat{\psi}_j$ \qquad from \quad $D_j$ \\
$\vdots$            &                              $\vdots$  \\
$D_K$ &\hspace{-2.4em} $\hat{\mu}_K$ \qquad from \qquad $D_1\ldots D_{K-1}$\\
      &\hspace{-6.6em} $\hat{\psi}_K$ \qquad from \qquad $D_K$\\
\end{tabular}
\end{center}
}}
\end{center}
\caption{In cross-fitting, the nuisance functions are estimated on part
of the data and the parameter is estimated on the held-out data.
This is done repeatedly and the final estimator is obtained
by averaging over the splits.}
\label{table::crossfitting}
\end{table}

We discuss two methods
for constructing confidence intervals:
cross-fitting and the HulC 
(Hull-based Confidence)
\citep{hulc}.
For cross-fitting with LOCO we use the following steps
\citep{williamson2020unified}.
\vspace{-.15in}
\begin{enumerate}
\item Divide the data into $K$ blocks
$D_1,\ldots, D_K$.
\item For block $j$, estimate $\mu$ from the data
in the other blocks and let 
$$
\hat\psi_j = \frac{1}{n_j}\sum_{i\in D_j} (\hat\mu(W_i,Z_i)-\hat\mu(Z_i))^2 \equiv
\frac{1}{n_j}\sum_{i\in D_j} T_{ij}
$$
where $n_j$ is the number of observations in $D_j$.
\item Let $\hat\psi = K^{-1}\sum_{j=1}^K \hat\psi_j$.
\item The confidence interval is
$\hat\psi \pm (z_{\alpha/2}\ s/\sqrt{n}+c/\sqrt{n})$ where
$s^2$ is the sample variance of the $T_{ij}$'s
and $c$ is the expansion constant from the previous section.
\end{enumerate}
\vspace{-.15in}
See Table \ref{table::crossfitting}.

An alternative is the HulC (see Table \ref{table::HulC}) which has
accurate coverage and does not require estimating the standard error
which is especially useful for the decorrelated LOCO
whose estimate is a $U$-statistic.
The steps are:
\vspace{-.2in}
\begin{enumerate}
\item Divide the data into
$B = \lceil\log(2/\alpha)/\log 2\rceil$ blocks.
\item
Split each block into two parts.
Estimate $\mu$ on the first part
and $\psi_j$ on the second part.
(We use a 4/5 and 1/5 split.)
\item 
The estimator is $\hat\psi = B^{-1}\sum_j \hat\psi_j$.
\item The confidence interval is\\
$[\min_j \hat\psi_j - c/\sqrt{n},\max_j \hat\psi_j + c/\sqrt{n}]$.
\end{enumerate}
\vspace{-.15in}

Cross-fitting has the advantage that it uses
more of the data to estimate the nuisance functions.
The advantage of the HulC is that we do not need to
estimate the standard error which, as we have mentioned,
is intractable for some parameters such as the
decorrelated LOCO.

\begin{table}
{\it The HulC}
\begin{center}
\fbox{\parbox{2in}{
\begin{center}
\begin{tabular}{lcc}

   &  \;\;$4/5$\quad             & \;$1/5$\quad\\
\\
$D_1$  &$\qquad\hat{\mu}_1\qquad$ & $\quad\hat{\psi}_1\quad$\\   
$D_2$  &$\qquad\hat{\mu}_2\qquad$ & $\quad\hat{\psi}_2\quad$\\
$\cdot$&         $\cdot$          &       $\cdot$          \\
$D_j$  &$\qquad\hat{\mu}_j\qquad$ & $\quad\hat{\psi}_j\quad$\\
$\cdot$&         $\cdot$          &       $\cdot$          \\
$D_B$  &$\qquad\hat{\mu}_B\qquad$ & $\quad\hat{\psi}_B\quad$\\
\end{tabular}
\end{center}
}}
\end{center}
\caption{With the HulC, splitting is done within each block of data.}
\label{table::HulC}
\end{table}

If we assume that $\mu$ is in a Holder space
of smoothness $\tau$,
then the validity of the interval
for LOCO requires that $\tau > d/2$.
If this assumption does not hold,
the confidence intervals will be biased
but can still be interpreted as a variability
interval, meaning that it covers
$\E[\hat\psi]$.
That is
$$
P(\E[\hat\psi]\in C_n)\to 1-\alpha.
$$
For the decorrelated LOCO
there is the possibility of additional bias
due to the fact that we did not use the influence function
and due to low density regions.
Again, these are probably best regarded as variability intervals.

For completeness,
we record the
efficient influence functions for
the correlation free parameters
\citep{verdinelli2021decorrelated}
in case the reader wants to bias correct the estimators.
For the decorrelated LOCO
the influence function is
\begin{align*}
\phi(X,Y,\mu,p) &=
\int[\mu(x,Z) - \mu_0(Z)]^2 p(x) dx \\
& +
\int[\mu(X,z) - \mu_0(z)]^2 p(z) dz\\
& \ +
2 \ \frac{p(X)p(Z)}{p(X,Z)}[\mu(X,Z)-\mu_0(Z)]\\
& \qquad
[Y - \mu(X,Z)] - 2\ \psi_{Dloco}(p).
\end{align*}

The influence function for $\psi_{cde}$ is
\begin{align*}
& \phi(X,Y,\mu,p) =
\int\int [\mu(X_1,z)-\mu(x_2,z)]^2 p(x_2)p(z)\\
& +
\int\int [\mu(x_1,z)-\mu(X_2,z)]^2 p(x_1)p(z) \\
& +
\int\int [\mu(x_1,Z)-\mu(x_2,Z)]^2 p(x_1)p(x_2)\\
& +
2\ \frac{Y - \mu(X_1,Z)}{p(X_1,Z)}  \int [\mu(X_1,Z)-\mu(x_2,Z)] p(x_2) d x_2 \\
& -
2\ \frac{Y - \mu(X_2,Z)}{p(X_2,Z)}\int [\mu(x_1,Z)-\mu(X_2,Z)]p(x_1) d x_1 \\& -3\ \psi_{cde}
\end{align*}

The lack of intrinsic second order accuracy
appears to be inevitable for decorrelated parameters.
Estimating these influence functions
is difficult and in particular, requires density estimation.
Doing so would destroy the simplicity of the procedure
and violate V1.

\subsection{LOCO for Forests}

\begin{figure}
\begin{center}
\includegraphics[scale=.3]{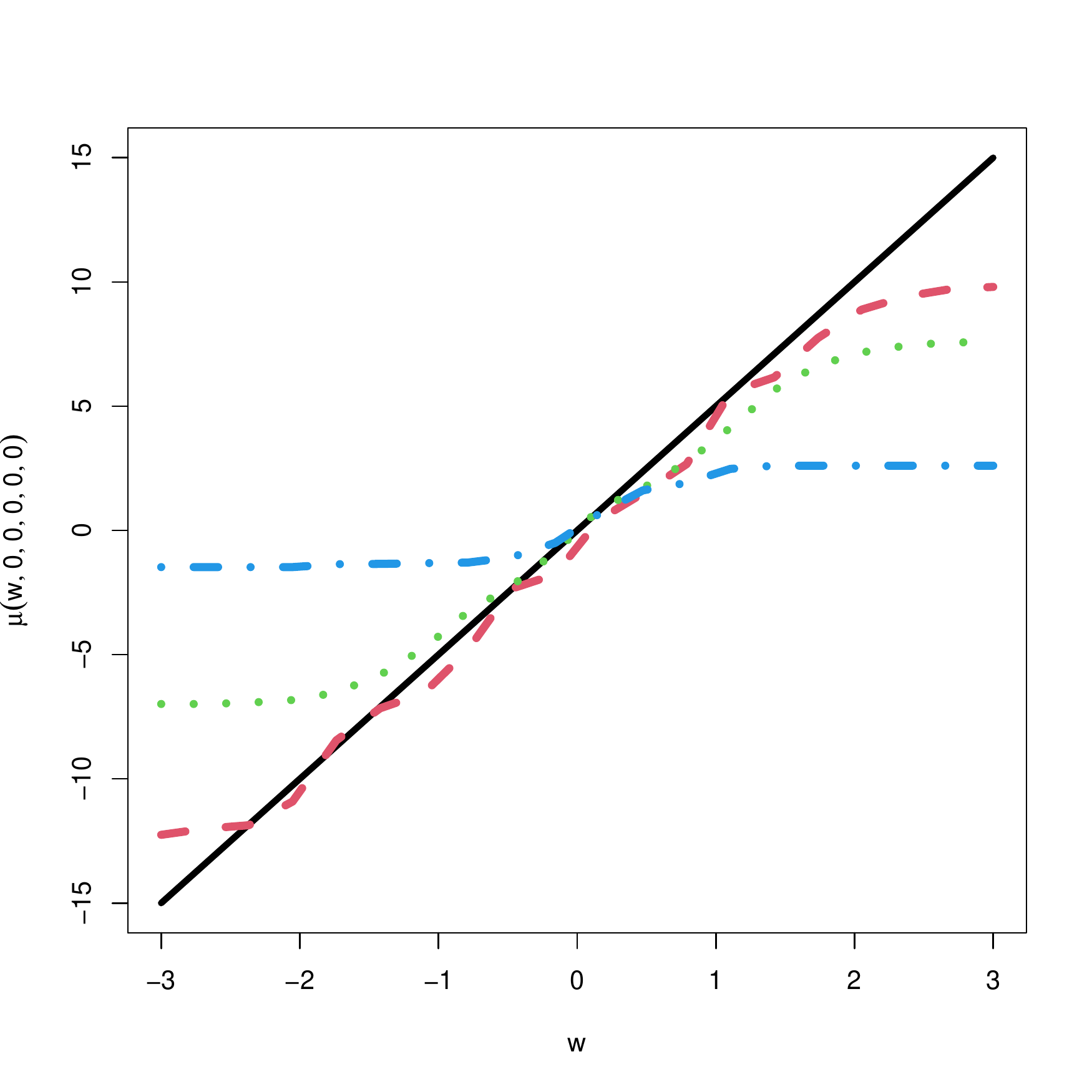}
\end{center}
\vspace{-.2in} 
\caption{\it True regression function 
$\mu(x,z)$ for $z$ fixed at $(0,\ldots, 0)$ 
(black solid line). The other curves are the estimates
from the random forest
with correlation $rho = .1$ (red), $rho = .5$ (green) and
$\rho = .9$ (blue). The forest estimate is very biased.}
\label{fig::problemN}
\end{figure}
 
\begin{figure}
\begin{center}
\includegraphics[scale=.3]{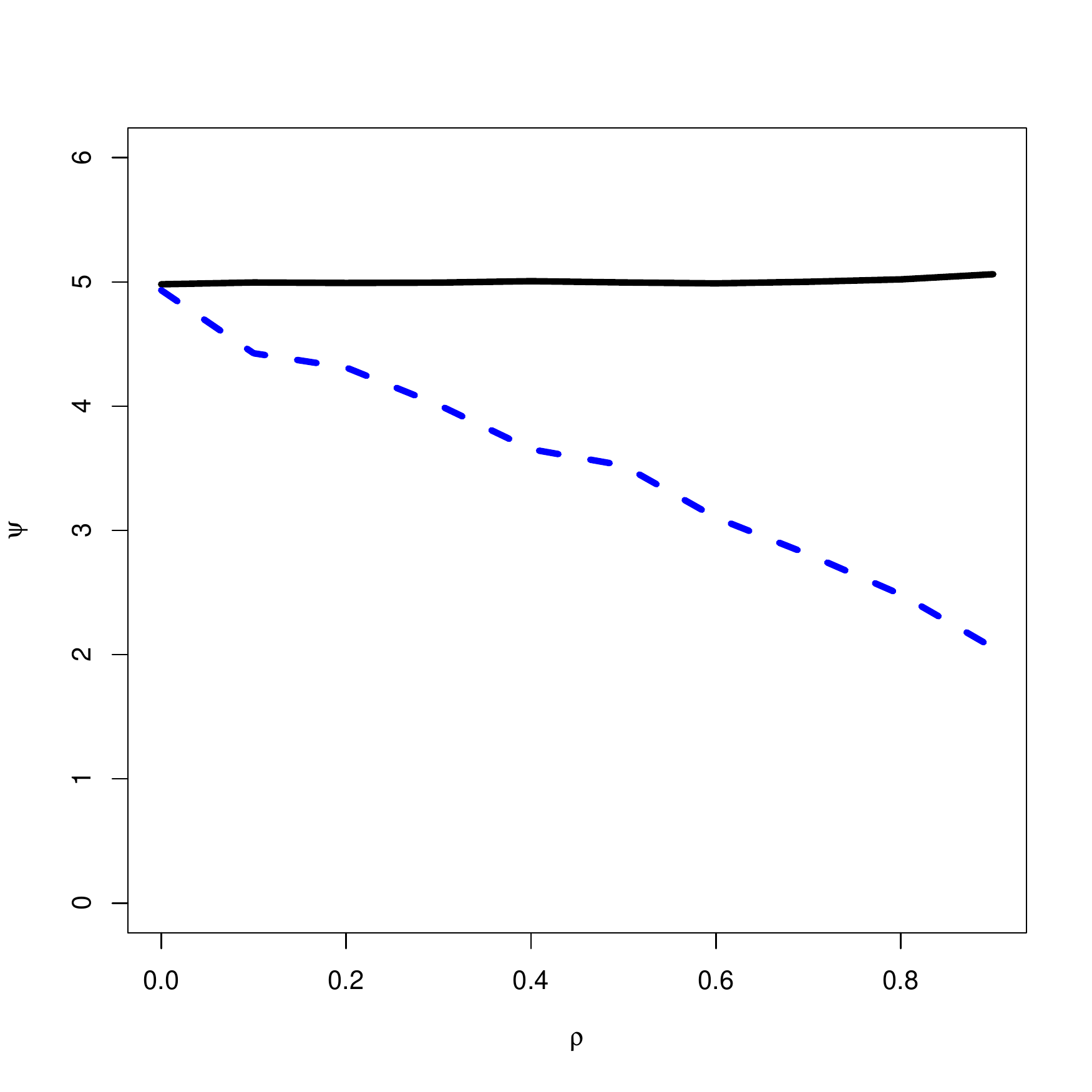}
\end{center}
\vspace{-.2in} 
\caption{\it The estimated normalized LOCO as a function of
$\rho$ using standard random forests (blue broken line) and 
the residual forest (black solid line). The standard forest 
is biased toward $0$.}
\label{fig::problem}
\end{figure}

Another issue arises when estimating LOCO
using
random forests.
We illustrate the problem
with the following example.
Let
$Y = \beta W + \gamma \sum_{s=1}^d Z_s + \epsilon$
where $\epsilon\sim N(0,1)$.
We take
$\beta = 5$, $\gamma = -2$,
$d=20$, $(W,Z)\sim N(0,\Sigma)$
where $\Sigma_{jj}=1$ and $\Sigma_{jk} = \rho$.
The normalized LOCO is $\beta^2$
but the estimate
$$
\hat\psi = 
\frac{\sum_i (\hat\mu(W_i,Z_i)-\hat\mu(Z_i))^2}{\sum_i (W_i - \hat\nu(Z_i))^2}
$$
using off-the-shelf random forests will be 
biased downwards due to overfitting.
(We used the R package \texttt{grf} but we verified that
the same problem happens with the packages \texttt{ranger} and
\texttt{randomForest}).
Figure \ref{fig::problemN} shows the estimated function
$\hat\mu(w,z)$ at several values of $w$ with $z=(0,0,0,0,0)$
using a random forest, for $\rho=0.1, 0.5$ and $0.9$.
The true function is $5w$
(black, solid line).
We see that the fitted functions are biased
as $\rho$ increases.
The plot in Figure \ref{fig::problem} shows the estimated LOCO
as a function of $\rho$ on the square root scale
(blue broken line).

The true value is 5.
The estimate
is biased downwards
with increasing bias as the correlation $\rho$ increases.

It seems reasonable to require that
an off-the-shelf method should have
good behavior when the model is linear or
approximately linear.
To achieve this we use a modified
forest as follows.
Let us write
$$
\mu(x) = L(x) + f(x)
$$
where
$L(x) = \theta^T x$ is the projection
of $\mu(x)$ onto the space of linear functions
and
$f(x)$ is the function orthogonal to $L(x)$.
That is, $\theta$ minimizes
$\E[(\mu(X)-\theta^T X)^2]$ and $f(x)$ lives in the orthogonal complement.
We can fit such a function easily
by first fitting a linear regression and then
fitting a forest (or other black box method)
to the residuals.
The black solid line in Figure \ref{fig::problem} shows 
the result in our example.
We refer to this method as a 
{\em residual forest}
and we use this method in all our examples.

\section{Examples}
\label{sec::examples}

In this section we present four simulated examples.
In each case, we plot the estimated parameters
(averaged over 100 simulations)
as a function of increasing correlation $\rho$ between the features.
In each case, the sample size is $n=1000$.

{\bf Example 1.}
Here the model is linear with
$Y = \beta W + \epsilon$,
$\epsilon \sim N(0,1)$,
$\beta = 5$,
$(W,Z)\sim N(0,\Sigma)$ is 10 dimensional,
$\Sigma_{jj}=1$,
$\Sigma_{12}=\Sigma_{21}=\rho$ and
$\Sigma_{ij}=0$ otherwise.

{\bf Example 2.}
The structure is the same as in Example 1 except
that $Y = 5 \cos(W) + 5 \cos(Z_1) + \epsilon$.

{\bf Example 3.}
This example is from
\cite{benard2021}.
Here, $Z$ is four dimensional,
$(W,Z)\sim N(0,\Sigma)$,
$\Sigma_{jj}=1$,
$\Sigma_{1,2} = \Sigma_{2,1} = \Sigma_{4,5} = \Sigma_{5,4} = \rho$
and
$$
Y = 1.5 W Z_1 I(Z_2 > 0) + Z_3 Z_4 I(Z_2 < 0) + \epsilon.
$$

{\bf Example 4.}
This example illustrates a distribution with a large region
of essentially 0 density.
We take $W\sim N(0,1)$,
$Z_1 = 3 W^2 + \delta$,
$(Z_2,\ldots,Z_9)$ are standard multivariate Normal,
$Y = 5 W + \epsilon$,
$\delta,\epsilon \sim N(0,1)$.

\begin{figure*}
\begin{center}
\includegraphics[scale=.8]{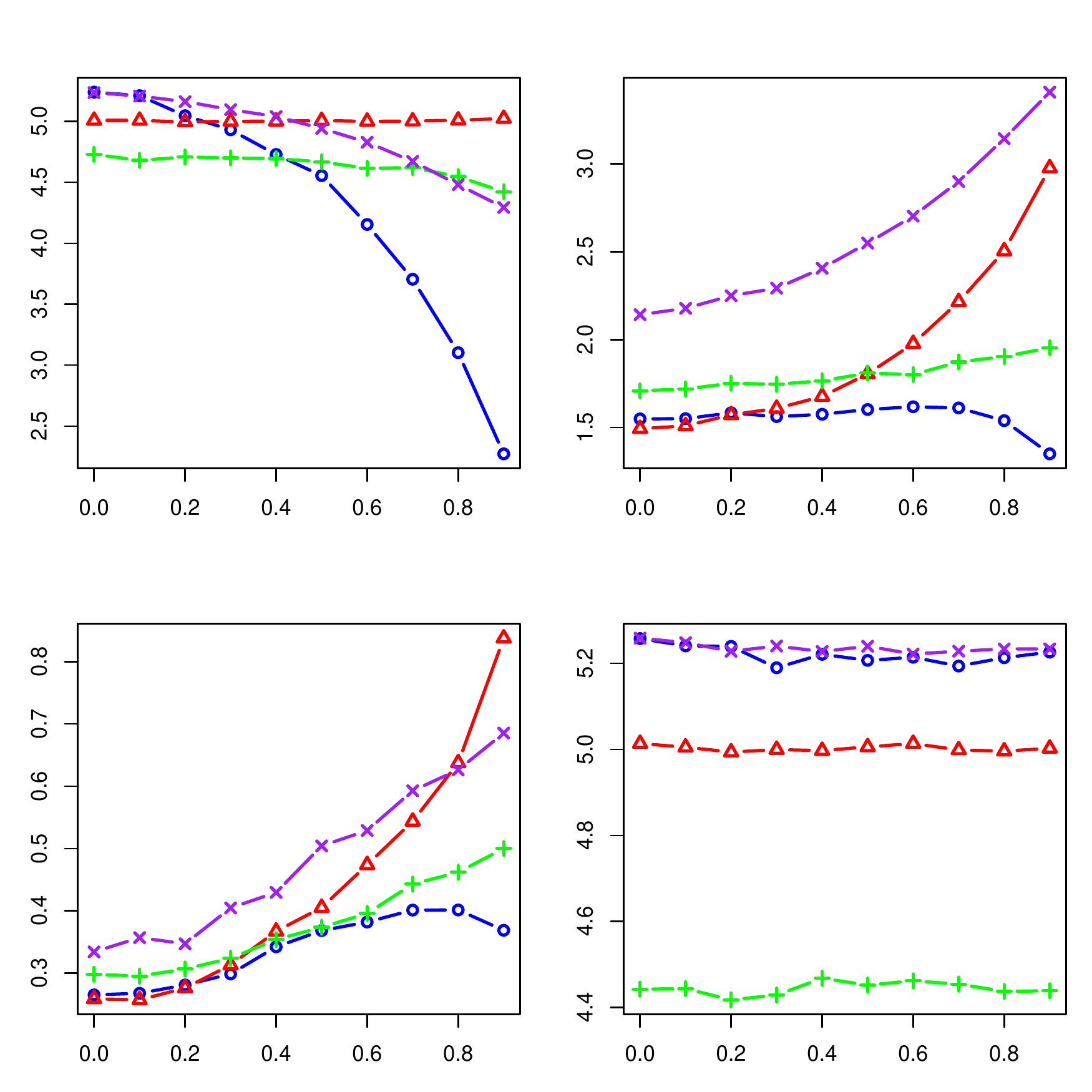}
\end{center}
\vspace{-.2in}
\caption{The four examples. Top left,  Example 1. Top right, Example 2. Bottom left, Example 3. Bottom right, Example 4. 
The blue curve {\bf \textcolor{blue}{$-\text{o}-$}} is the unnormalized LOCO.
The red curve {\bf\textcolor{red}{$-\triangle-$}} is the normalized
LOCO. The purple curve {\bf\textcolor{purple}{$-\times-$}} is the Shapley value.
The green curve \textcolor{green}{$-+-$} is the decorrelated LOCO.
The horizontal axis is the correlation $\rho$.
The effect of feature correlation is clear especially for the unnormalized
LOCO and the Shapley value. In the fourth example (bottom right plot)
the decorrelated LOCO is badly biased due to low density bias.}
\label{fig::examples}
\end{figure*}

From Figure \ref{fig::examples}
we clearly see the effect of correlation.
$\psi_{Dloco}$ is least affected by correlation as expected
but in the bottom right plot it is too small due to low density bias.
The Shapley value is affected by correlation
and in some cases, it moves in the opposite direction of LOCO
(top right and bottom left).
Most of the time, the normalized LOCO does better than the
Shapley value.
One might have hoped that the Shapley value would lie in between
$\psi_{Dloco}$ and the unnormalized LOCO.
This might be achieved using a modified Shapley value
as described in Section \ref{sec::modified}.

\begin{figure}
\begin{center}
\includegraphics[scale=.5]{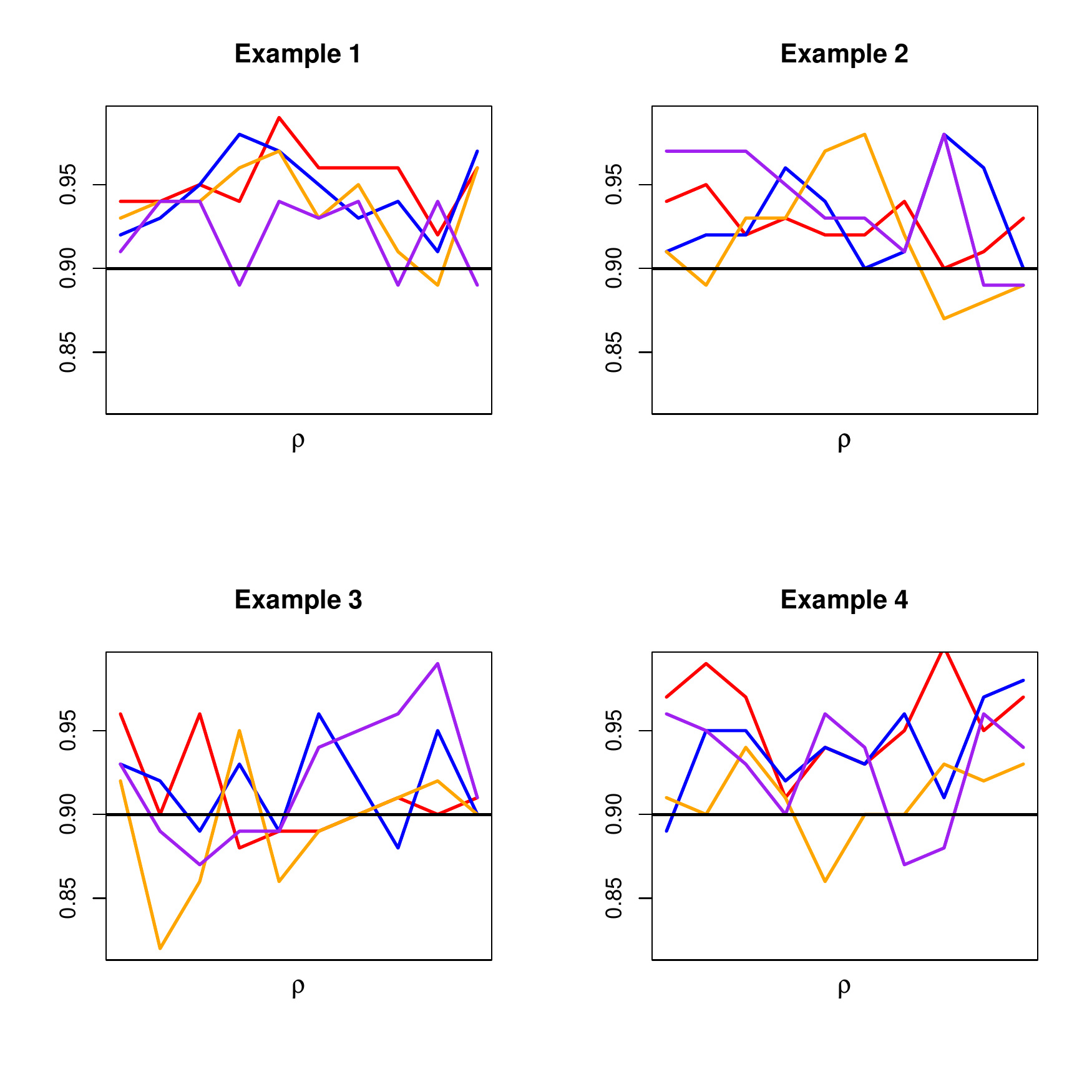}
\end{center}
\vspace{-.3in}
\caption{The plots show the coverage of the variability intervals\\
using the HulC for the four examples. The nominal level is 0.9.\\
Red: LOCO, Blue: Dloco, Orange: Unnormalized LOCO, Purple:\\
Shapley. In most cases the coverage is close to nominal for all\\
values of the correlation $\rho$.}
\label{fig::coverage}
\end{figure}

We also conducted a small simulation
to check the coverage of the 90 percent variability intervals
using the HulC. The results are shown in Figure \ref{fig::coverage}.
In most cases the coverages are close to nominal or higher.

\section{Groups of Variables}
\label{sec::groups}

So far we have assumed that
the feature of interest $W$ is scalar.
More generally, we can consider a group of variables
$W\in\R^g$.
LOCO is defined in the same way, that is,
$$
\psi_{loco}(W) = \E[ (\mu(W,Z)-\mu(Z))^2]
$$
where again $Z$ denotes the rest of the features.
In fact, looking at groups
of features is another way to deal
with correlation.
Suppose we are interested in a feature $X_j$.
If the single feature of interest $X_j$ is highly correlated
with some other features $F$
then it might be useful to compute LOCO
for $W = (X_j,F)$.
In fact, we might consider nested subsets
$W_1\subset W_2 \subset \cdots$ of decreasing correlation
each containing $X_j$
as a way of exploring the importance of $X_j$.
We can form these groups
in a forward stepwise fashion as follows:

\begin{figure}
\begin{center}
\includegraphics[scale=.4]{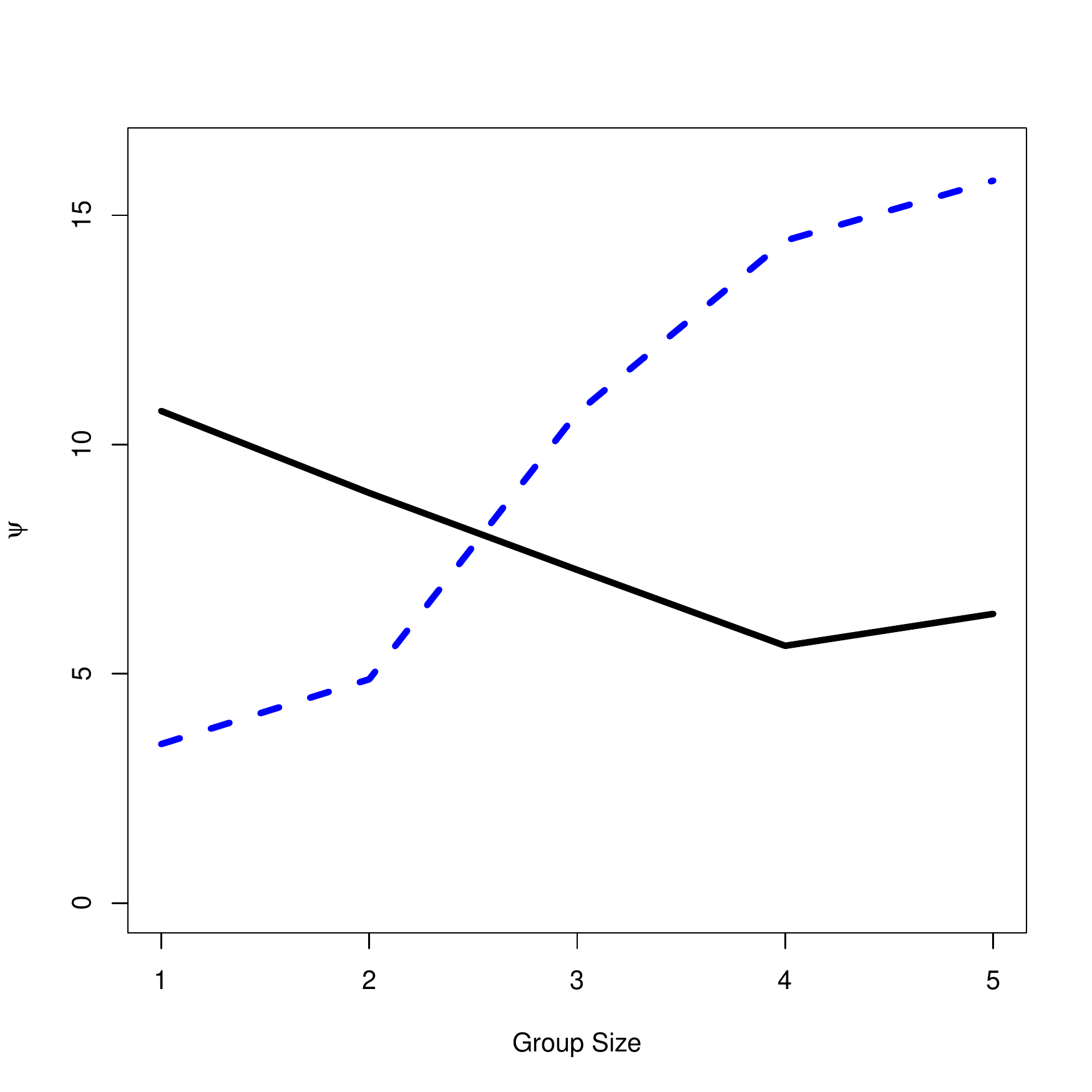} 
\vspace{-.2in} 
\caption{\it An example showing the un-normalized LOCO (blue, broken line)
and normalized LOCO (black, solid line) as the group size is grown using the
forward selection process described in the text.
The un-normalized LOCO has correlation distortion that decreases
as the group is grown. This value necessarily increases with group
size (apart from some random variability involved in the estimation
procedure). The normalized LOCO is less sensitive
to correlation distortion but does not have monotone behavior in the
group size.}
\label{fig::groups}
\end{center}
\end{figure}

\begin{enumerate}
\item Set $W = X_j$.
\item Compute $\psi_{loco}(W)$.
\item Find the $Z_j$ with highest absolute correlation with $W$.
Add this $Z_j$ to $W$ and compute $\psi_{loco}(W)$.
\item Continue until all variables are included in $W$.
\item Plot $\psi_{loco}$ as a function of the number of
variables in $W$.
\end{enumerate}

Any measure of correlation can be used.
We used canonical correlation  
to explore the groups of features behavior in Figure \ref{fig::groups}.
We considered an example where $n=1000$, $(W,Z)\sim N(0,\Sigma)$,
$\Sigma_{jj}=1$, 
$\Sigma_{1,2}=\Sigma_{1,3}=\Sigma_{2,1}=\Sigma_{3,1}=
\Sigma_{2,3}=\Sigma_{3,2}=.95$
and $\Sigma_{j,k}=0$ otherwise.
The dimension of $Z$ is 9 and
$Y = 2W^2 Z_1^2 + \epsilon$.
One can either use the original LOCO
or the normalized version.
Recall that the normalized version
was designed to recover the least squares
coefficient in linear models dividing by
$\E[(W-\nu(Z))^2]$
where $\nu(z) = \E[W|Z=z]$.
When $W$ is a vector this term becomes a matrix, namely,
$\E[(W-\nu(Z)) (W-\nu(Z))^T]$.
In our example we normalized by the trace of this matrix,
but other choices are possible.
But recovering $||\beta||^2$ is not possible.
The unnormalized LOCO has the advantage that it is non-decreasing
as the group is grown which may be considered 
an advantage for interpretation.

We see that the unnormalized LOCO (blue, broken line)
suffers correlation distortion that becomes less severe
as the group is grown.
The normalized LOCO (black, solid line) suffers less correlation distortion
but does not have the intuitive monotone behavior that
the unnormalized LOCO has.

Extending Shapley values for groups
is straightforward.
If the variables
are put into groups
$G_1,\ldots, G_k$
we can now use
$$
\psi_{shap}(G_j) = \sum_S w_S \ \psi_{loco}(G_j,S)
$$
where now the sum is over subsets of groups.

\section{Modified Shapley Values}
\label{sec::modified}

Although Shapley values do not fix correlation distortion,
they do sometimes reduce it.
For instance, in Case 2 of the example in Section
\ref{sec::correlation},
we had, for the unnormalized LOCO that
$\psi_{loco} =0$ while
$\psi_{shap} = \beta^2/d$
which provides at least some improvement.
(We focus on the un-normalized LOCO in this section.)
We believe that this is why some authors claim
that Shapley values address correlation.
Apparently, averaging over submodels can, in some cases,
partially reduce correlation distortion.

Suppose that 
$X = (W,Z)$ and that
there is a subset
of features $S$
such that 
(i) $\mu(w,z) = \mu(w,x_S)$ and such that
(ii) $W$ is independent of $X_S$.
In that case, it is easy to see that
$\psi_{loco}(W,S)$
(the LOCO value for $W$ in the model based on $(W, X_S)$)
is equal to $\psi_{Dloco}$.
Finding a submodel $S$ with exactly these properties
is unlikely.
But if we modify the Shapley
value to give high weight to submodels 
that approximately satisfy (i) and (ii)
we can get an estimate that is closer to
$\psi_{Dloco}$ than the usual Shapley value.
Making this idea precise could be a topic
for a whole paper.
Here, we give one example
of how one might define such a modified Shapley value,
as a proof of concept.
(See \cite{mase2022} for further discussion on this topic.)

\begin{figure}
\begin{center}
\begin{tabular}{cc}
\includegraphics[scale=.2]{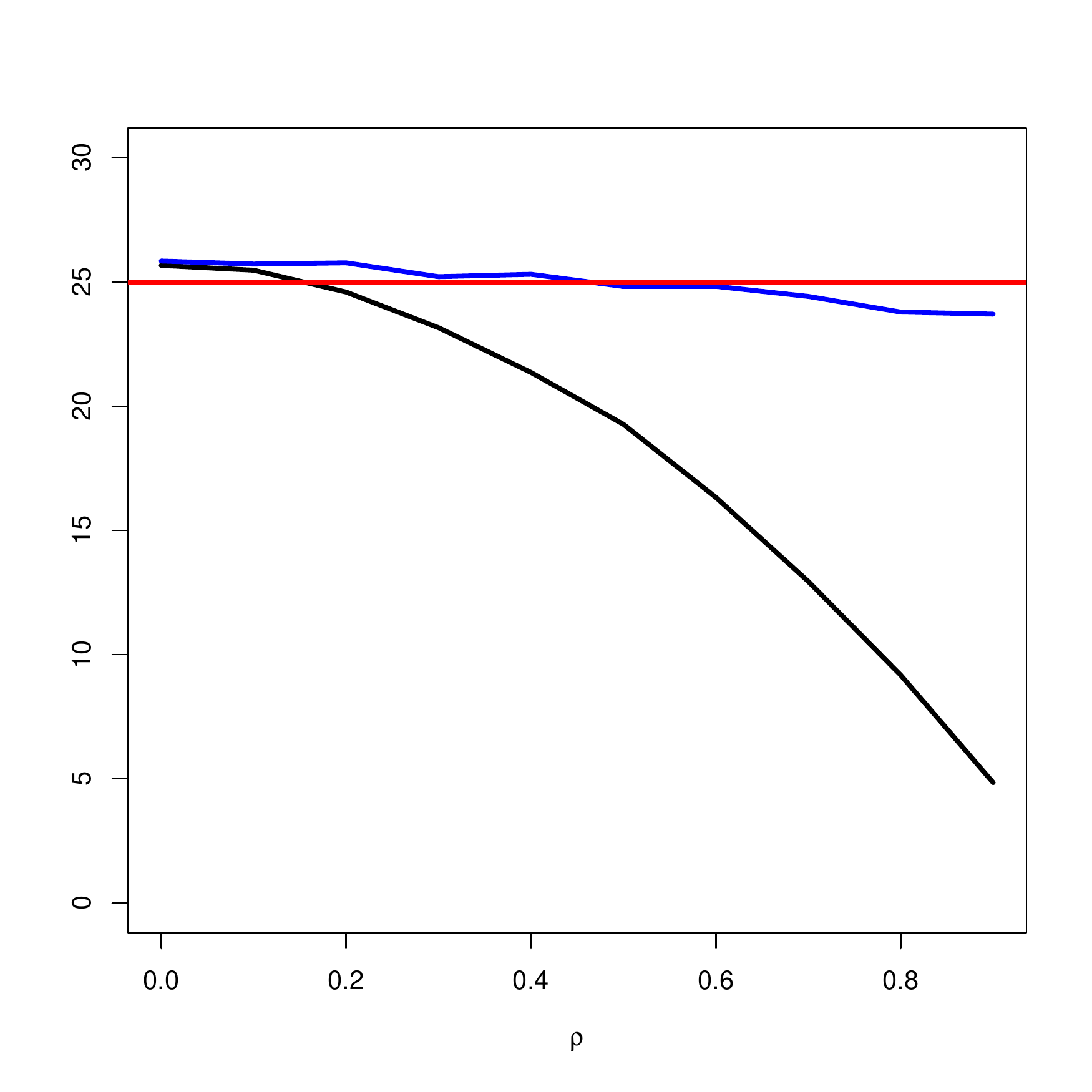} &
\includegraphics[scale=.2]{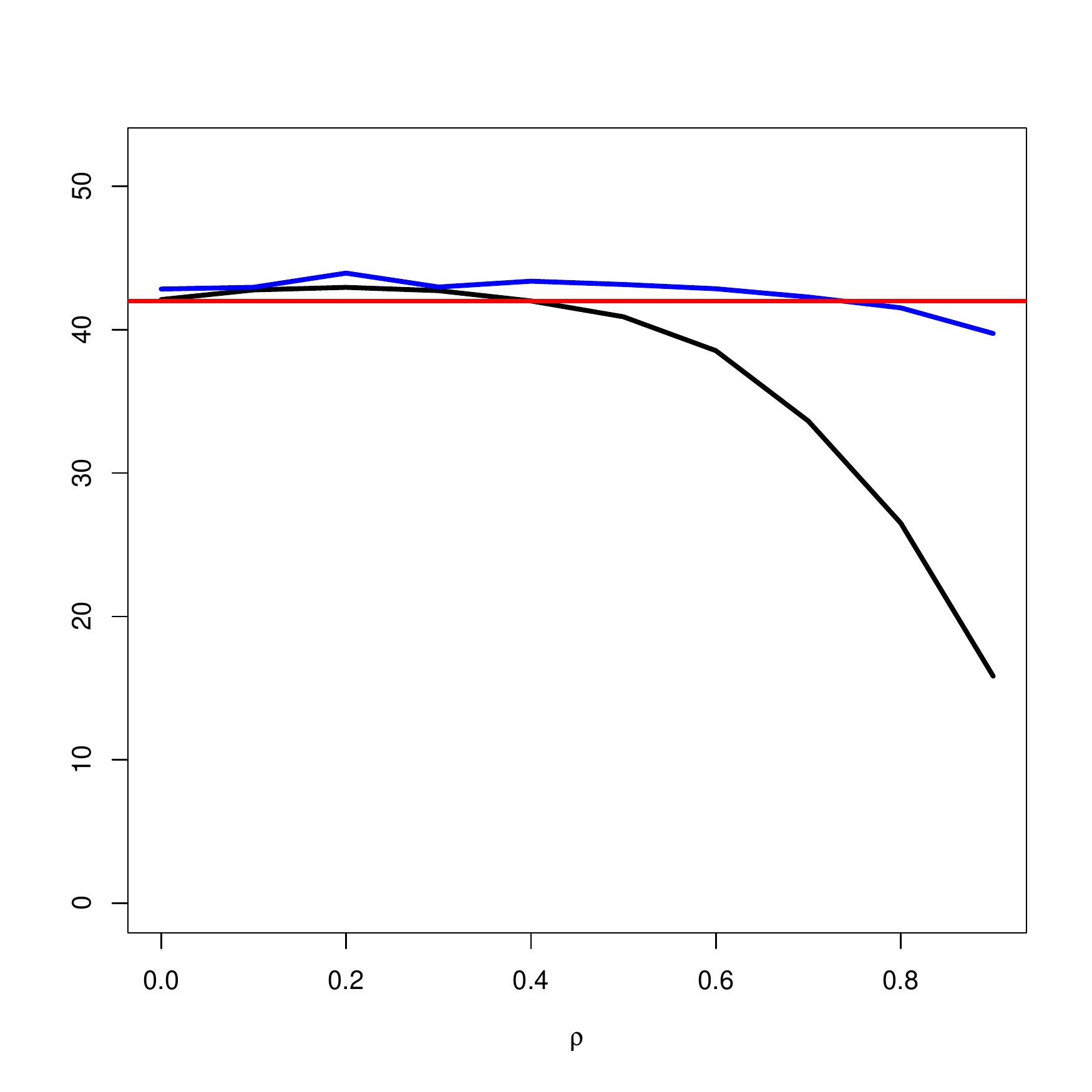}
\end{tabular}
\vspace{-.1in} 
\caption{\it Left panel: The linear model of Section \ref{sec::correlation} 
revisited. Right panel: nonlinear model with $Y = 5X^2 + \epsilon$
and equal correlation $\rho$ between all covariates.
The decorrelated LOCO (red line), the
un-normalized LOCO (black line) and the modified Shapley
value (blue line) for several values of $\rho$.}
\label{fig::modified}
\end{center}
\vspace{-.2in}
\end{figure}

We want to give high weight to any model
$S$ that is an accurate predictor and such that
$W$ is independent of $(Z_s:\ s\in S)$.
Define weights
$$
w_S \propto \frac{\text{acc}}{\text{dep}}
$$
where $\text{acc}$ is some measure of the accuracy
of the submodel $(W,Z_S)$ and
$\text{dep}$ is a measure of the dependence between
$W$ and $Z_S$.
As a specific case, we can take
$\text{acc} = 1/\psi_{loco}(S^c)$
and $\text{dep}$ to be the canonical correlation
between $W$ and $S$ denoted by $\rho(W,S)$.
Hence we define
$$
w_S \propto \frac{1}{\psi_{loco}(S^c) \rho(W,S)}.
$$
As with the usual Shapley values
we use subsampling to compute the modified Shapley value.

Two examples
are shown in Figure
\ref{fig::modified}.
The left plot is the linear model
from Section \ref{sec::correlation}.
The right plot is a nonlinear example
with $Y = 5X^2 + \epsilon$
and equal correlation $\rho$ between all covariates.
In each case, the horizontal red line is $\psi_{Dloco}$,
the black line is LOCO and the blue line is the modified Shapley value.
The latter better approximates $\psi_{Dloco}$ for all values of $\rho$.
A full exploration of this idea
is beyond the scope of this paper,
but it is important to be aware
that we do not have to feel compelled
to use the standard definition of Shapley values.
Further investigation into modified versions of Shapley
values is an important priority.

\section{Conclusions and Recommendations}
\label{sec::conclusion}

Table \ref{table::summary}
summarizes properties of some variable importance
measures. 

{\bf LOCO}.
The normalized version of LOCO
is easy to compute, has second order bias,
does not suffer low density bias and is interpretable.
In linear models, it is correlation free but in general it does suffer
from correlation distortion which, as we have seen can deflate or inflate
its value.

{\bf Decorrelated LOCO.}
This parameter is free from correlation distortion.
But it suffers first order bias. To correct this bias,
we need to estimate the influence function which entails
density estimation. This parameter is subject to low density bias
which does not occur for LOCO.

{\bf Shapley.}
The main attraction of the Shapley value is
its axiomatic basis.
But as we have pointed out, outside of its original game theory setting,
the axioms are questionable.
The Shapley value is very expensive to compute, is difficult to interpret
and still suffers correlation distortion.
There may be specialized circumstances where Shapley values
are useful --- such as computer experiments where the distribution of
$X$ is under the control of the user,
but we do not recommend them for general use.

{\bf Variability Intervals versus Confidence Intervals.}
In many applications it may suffice to only have point estimates.
But if the user wants a confidence interval,
we have seen that these can be obtained either
through cross-fitting or the HulC.
These intervals are best interpreted as variability intervals not
confidence intervals
due to the estimation bias.

\begin{table}
\caption{\it A comparison of (normalized) LOCO, decorrelated LOCO, and Shapley. Each parameter has advantages and disadvantages. Overall, LOCO seems the best choice.}
\label{table::summary}
\begin{center}
\fbox{\parbox{3in}{
\begin{tabular}{lccc}
                      &   LOCO       & dLOCO     & Shapley  \\
Correlation free      &   $\times$   & $\surd$   & $\times$ \\
Second order bias     &   $\surd$    & $\times$  & $\surd$  \\
No low density bias   &   $\surd$    & $\times$  & $\surd$  \\
Fast                  &   $\surd$    & $\surd$   & $\times$ \\
Interpretable         &   $\surd$    & $\surd$   & $\times$ \\
\end{tabular}
}}
\end{center}
\end{table}

{\bf Modified Shapley Values.}
The real benefit of Shapley values is that they explore
submodels which might have useful information.
The standard definition of the Shapley weights comes from
cooperative game theory.
But as we noted in Section \ref{sec::modified}
we can easily change the definition to suit our purpose.
This opens up a wealth of possibilities that should be explored.

{\bf Other Value Functions.}
Although we focused on a particular value function $V(S)$,
\cite{kumar2020problems} show that the same issues arise
with Shapley values when using other value functions.
They consider two different value functions:
the conditional mean
$V(S) = \E[Y|X_S=x_S]$
and the intervention distribution
$V'(S) = \int \mu(x_S,x_{S^c}) dP(x_{S^c})$.
The latter is the mean of the counterfactual value of
$Y$ if we intervened and set $X_S$ to equal the specific value of $x_S$.
However, caution is warranted
since this is a causal quantity and the interpretation
that this measures an intervention depends on
a variety of assumptions, especially the assumption of no unmeasured
confounding. 
The authors say:

{\em ... we demonstrate that Shapley-value-based explanations
for feature importance fail to serve their desired purpose in
general. ... applying the Shapley value to the problem of feature importance
introduces mathematically formalizable properties which may not
align with what we would expect from an explanation.}

They point out that
when using $V(S)$,
the Shapley value suffers correlation distortion.
They give an extreme example, namely, including or excluding
a feature that is perfectly correlated with another feature
does not change the value of $\mu(x)$ but it does change the Shapley value.
(This corresponds to a violation of our axiom A1.)
And when using $V'(S)$,
we get extrapolation bias due to regions in the sample space with low density.
(This corresponds to a violation of our axiom V3.)

\bigskip

Finally, as we mentioned in the introduction,
there are many other feature importance measures 
in the statistics and machine learning literature.
We focused on LOCO and Shapley because of their
popularity.


\bibliographystyle{imsart-nameyear} 

\bibliography{paper}       

\end{document}